\documentstyle[psfig,aps,prl,twocolumn]{revtex}
\newcommand{\be}{\begin{equation}}
\newcommand{\ee}{\end{equation}}
\newcommand{\bea}{\begin{eqnarray}}
\newcommand{\eea}{\end{eqnarray}}

\newcommand{\bfo}{\mbox{\boldmath $\omega$}}
\newcommand{\bfd}{\mbox{\boldmath $\delta$}}
\newcommand{\bft}{\mbox{\boldmath $\tau$}}

\begin{document}
\twocolumn[
\begin{center}
{\LARGE\bf\sf Ray and wave chaos in asymmetric resonant optical cavities}\\
\vspace*{0.5cm}
{\large\sf 
Jens U.~N{\"o}ckel \& A.~Douglas Stone}\\
\vspace{0.5cm}
{\sf
Department of Applied Physics, Yale University, New
Haven, Connecticut 06520-8284, USA\\}
\vspace{0.5cm}
{{\em Published in} Nature {\bf 385}, 45 (1997)}
\end{center}
\vspace{1cm}
{\bf Optical resonators are essential components of
lasers and other wavelength-sensitive optical devices.
A resonator is characterized by a set of modes, each with a resonant
frequency $\bfo$ and resonance width $\bfd \bfo$ = 1$/\bft$, 
where $\bft$ is the lifetime of a photon in the mode.
In a cylindrical or spherical dielectric resonator, extremely 
long-lived resonances \cite{intro} are due to `whispering gallery' 
modes in which light circulates around the perimeter trapped by total
internal reflection.  These resonators emit light isotropically.
Recently, a new category of asymmetric resonant cavities (ARCs) has
been proposed in which substantial shape deformation leads to
partially chaotic ray dynamics. This has been predicted 
\cite{wgopt,prl,chapter} to give rise to a universal, frequency-independent
broadening of the whispering-gallery resonances, 
and highly anisotropic emission.
Here we present solutions of the wave equation for ARCs
which confirm many aspects of the earlier ray-optics model, but 
also reveal interesting frequency-dependent effects
characteristic of quantum chaos.  For small deformations the 
lifetime is controlled by evanescent leakage,
the optical analogue of quantum tunneling \cite{Johnson}.
We find that the lifetime is much shortened
by a process known as `chaos-assisted tunneling' \cite{Bohigas,Doron}.
In contrast, for large deformations ($\sim\,$10\%) some resonances are found 
to have longer lifetimes than predicted by the ray chaos model due to
``dynamical localization'' \cite{Casati}.
}\\\vspace{1cm}
]

\flushbottom
The prediction of universal behavior of the whispering gallery resonances
is derived from the limit of ray optics in which
the wavelength of the light, $\lambda$, is much shorter than
the radius of curvature of the ARC.  Here for simplicity 
we will focus on the effectively two-dimensional 
case of a deformed cylindrical resonator as in Fig.~1 (top); the
case of deformed spheres (e.g. liquid droplets) 
in the ray limit has been discussed elsewhere
\cite{prl}.  When the cylinder is undeformed (circular), the angle
of incidence $\chi$ is conserved at each collision and escape occurs
isotropically by the exponentially slow process of evanescent leakage. 
When the cylinder is sufficiently deformed a new decay process,
refractive escape, becomes possible \cite{wgopt,chapter} in which
a ray initially in a whispering gallery trajectory diffuses chaotically
in phase space until it reaches the critical angle, $\chi_c = \sin^{-1}(1/n)$,
(where $n$ is the refractive index of the dielectric) and is refracted out
of the resonator.  This is illustrated schematically
in Fig.~1 (bottom).  Since no refractive escape is allowed for $\chi > \chi_c$,
the ray dynamics is equivalent to the non-linear dynamics of a point
mass undergoing specular reflection from the walls of a two-dimensional
``billiard'' \cite{Berry}.  For smooth convex deformations 
from a circle this dynamics
becomes increasingly chaotic with increasing deformation 
according to the KAM (Kolmogorov-Arnold-Moser) scenario  
\cite{chapter,Lazutkin,Robnik}.  One stage in this evolution is shown
in the Poincar\'e surface of section (Fig.~2a). 
Initially, small regions of chaotic behavior
appear near unstable periodic orbits which al-\\

\hbox{\psfig{figure=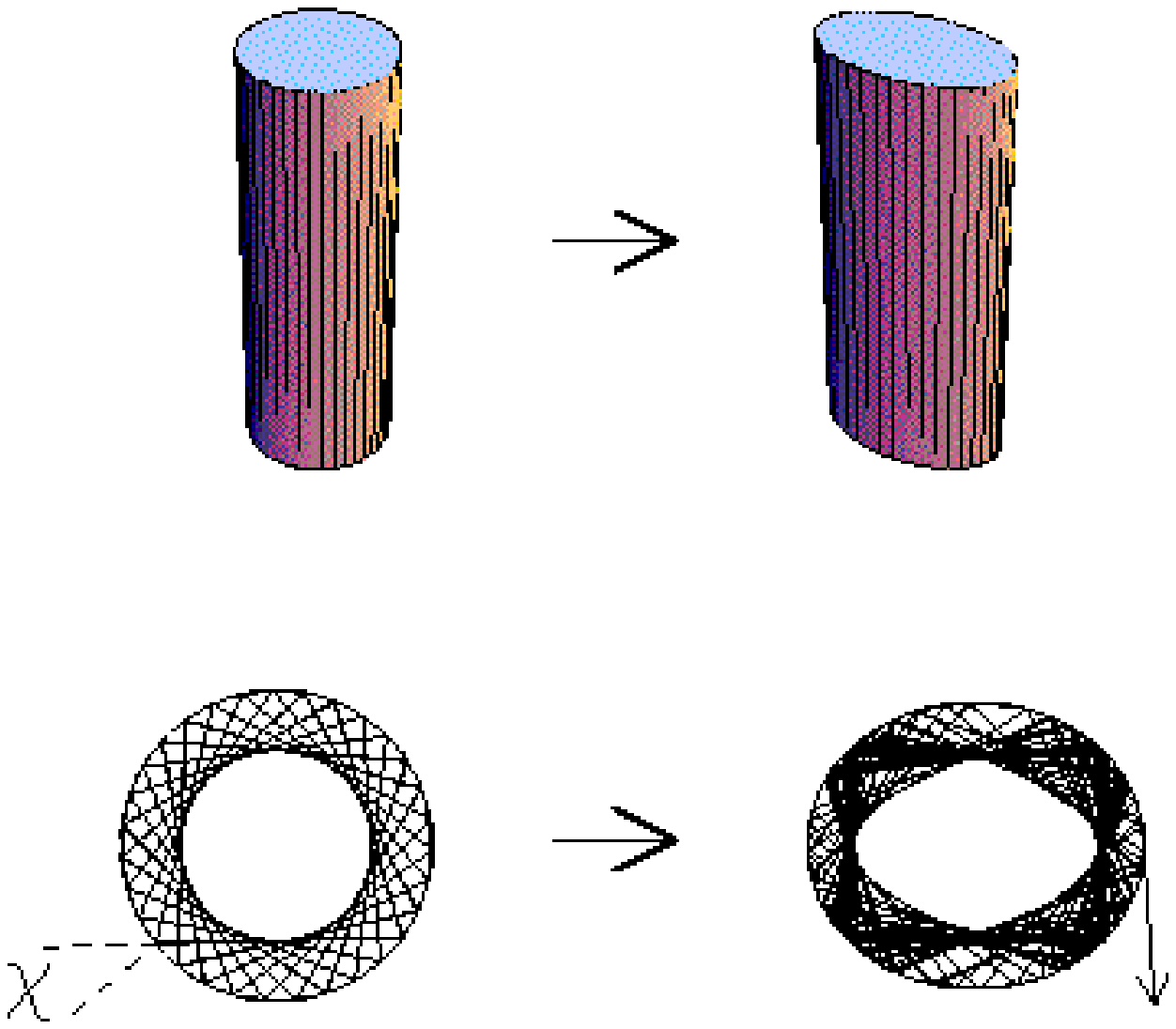,width=7.7cm}}
\footnotesize\noindent FIG.~1 Top, deformation of a dielectric cylinder
of radius $R$ to make an asymmetric resonant cavity (ARC).
All results below are for a
quadrupolar deformation of the cross-section, so that the
radius as a function of angle $r(\phi)= r_0(1+ \epsilon \cos
2 \phi)$(where $\epsilon < 0.2$ measures the size of 
the deformation and $r_0(\epsilon)$ is chosen to 
maintain a fixed area). 
Bottom, regular and chaotic ray trajectories in the plane perpendicular to the 
cylinder axis for a whispering gallery orbit for $\epsilon=0,\epsilon =0.1$; 
the chaotic trajectory eventually escapes refractively when 
the sine of the angle of incidence, $\sin \chi < 1/n$.

\twocolumn[
\vspace*{-2cm}
\hbox{\psfig{figure=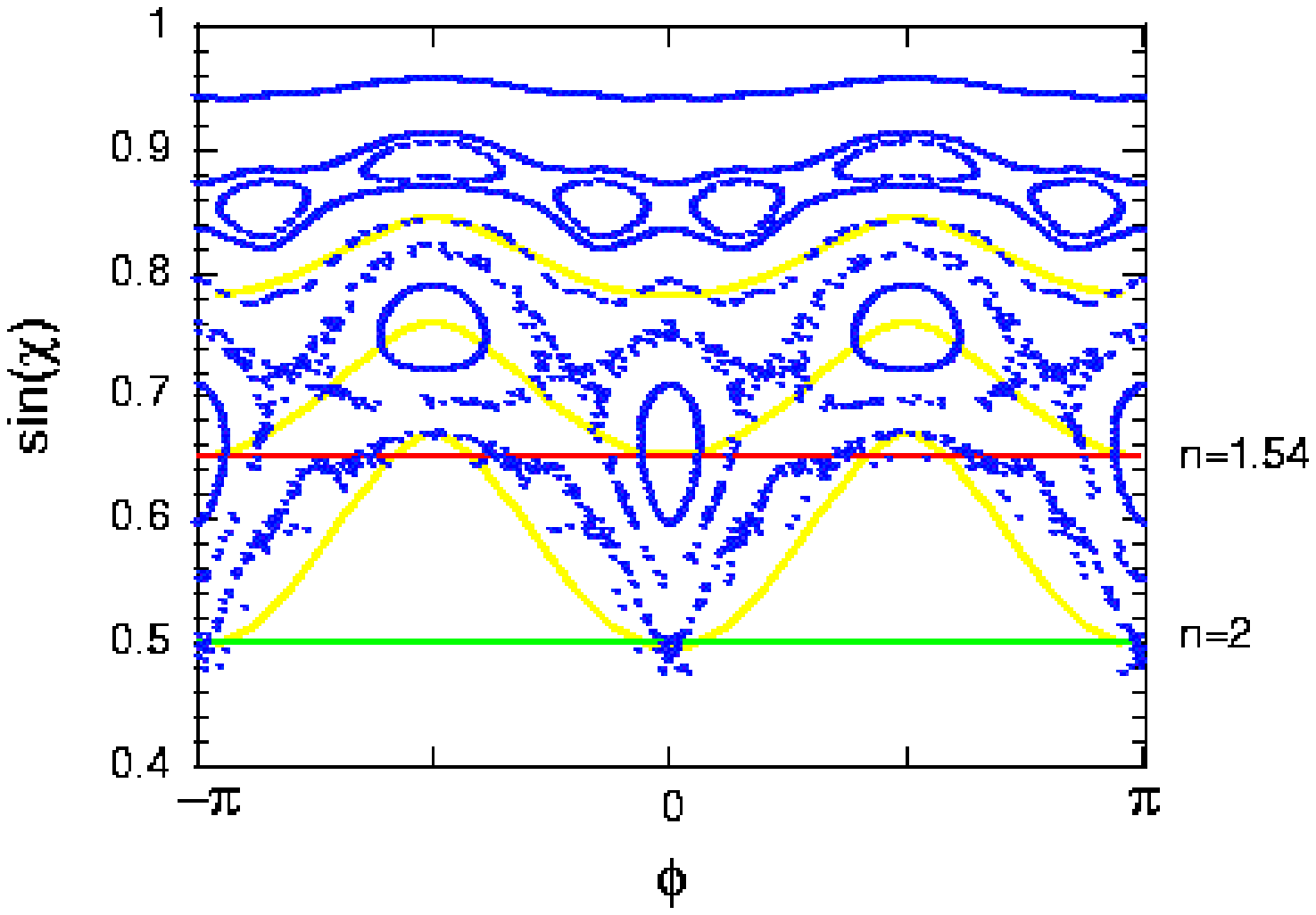,width=9.3cm}
\psfig{figure=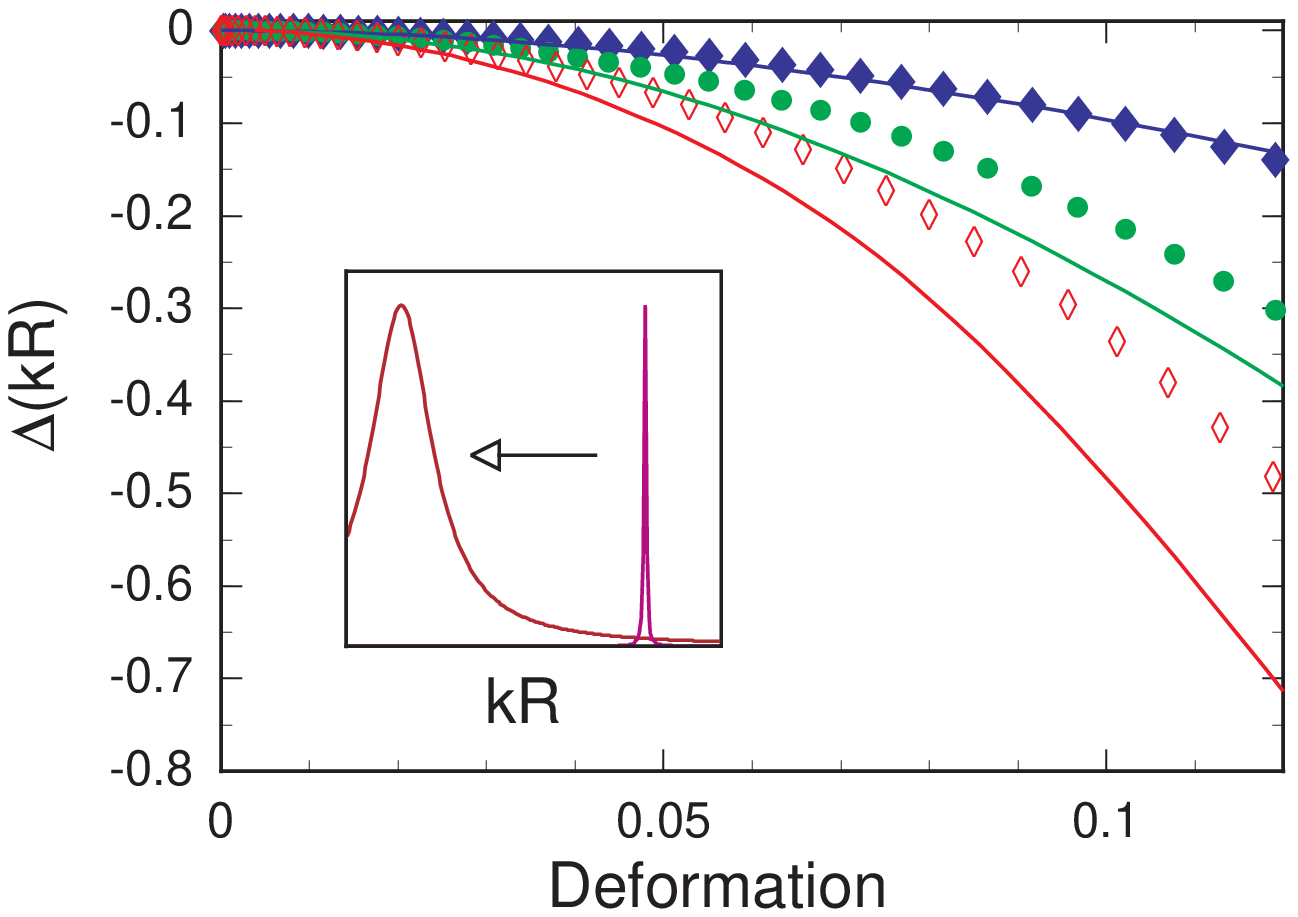,width=9.2cm}}
\vspace*{2mm}

\noindent
FIG.~2 {\it a}, 
Partial Poincar\'e surface of section showing phase space
ray dynamics for several trajectories for $\epsilon=0.072$.
The $\sin \chi$ and the angular position
$\phi$ are recorded at each collision.
The green and red horizontal lines represent the critical lines $\sin\chi=1/n$
for refractive escape for $n=2,1.54$.  ``Islands'' indicate
trajectories oscillating around stable periodic orbits.
Unbroken KAM curves appear above $\sin \chi \approx 0.8$; 
no whispering gallery orbit
can cross such a curve.  Trajectories below $\sin \chi \approx 0.8$
lying outside of islands are all chaotic, but when they are
followed for 100-200 reflections (as above) they remain near the
the adiabatic KAM curves predicted by Eq.~(1) (yellow);
except when islands intersect
the relevant curve (see the discusssion of peak-splitting
in Fig.~4 legend). 
{\it b}, Shift in resonance frequency with deformation from exact
numerical solution of the wave equation for the ARC resonances
(solid lines) and from eikonal approximation for the bound states
(symbols, adjusted to agree at $\epsilon =0$).  
Units of frequency are $kR = 2 \pi R/
\lambda$ where $\lambda$ is the wavelength outside the dielectric. For
$\epsilon=0$, the resonances are at $kR \approx 12.1,27.5,44.6$ (blue,green,
red).  Inset schematic shows the generic redshift and broadening of
these resonances with deformations.  The redshift is easily understood:
the approximate condition for a whispering gallery resonance 
is that an integer number
of wavelengths fit around the perimeter (see Fig. 4a).  As the ARC
is deformed for fixed area the perimeter increases requiring the
wavelength to increase.
\vspace*{1cm}]

\normalsize\noindent ternate with 
stable orbits and their
associated islands, but any chaotic trajectory is confined to a small
separatrix region near the islands by KAM curves
(which correspond to families of regular quasi-periodic orbits).
As the deformation increases these KAM curves break up, allowing 
refractive escape of whispering gallery orbits.
Because ray optics describes the limit $\lambda \to 0$, the
escape rate due to this process is independent of $\lambda$
and should be the same for all resonances corresponding to the same
set of whispering gallery orbits.

To compare the previous model derived from ray optics 
\cite{wgopt,chapter} to solutions of the wave
equation we have to overcome a fundamental problem.  
In regular (integrable) systems a set of trajectories on a KAM curve
corresponds to a set of quantized solutions of the wave equation;
these solutions are obtained by finding the KAM curves for which
the conserved actions are appropriately quantized \cite{Gutzwiller}.
For general chaotic systems there
exists no such correspondence \cite{Gutzwiller}.
For ARC resonances we propose here a method for establishing an approximate
correspondence.  It is known for 2D convex billiards that 
the regular orbits which follow KAM curves and generate caustics in the 
real-space ray motion are well-described by an adiabatic approximation 
\cite{AC}.  
The (adiabatic) KAM curves representing 
whispering gallery trajectories in the
surface of section of Fig.~2{\it a} are given by: 
\be\label{adiab}
\sin\chi(\phi)=\sqrt{1-\left(1-S^2\right)\kappa(\phi)^{2/3}}.
\ee
where $0<S<1$ parameterizes the adiabatic curve and $\kappa (\phi)$
is the curvature
of the ARC boundary at an azimuthal angle $\phi$.
Although for the deformations of interest the relevant adiabatic 
curves are ``broken''
and the long-time behavior of a chaotic whispering gallery 
orbit departs strongly
from Eq.~(1), we find that for intermediate times ($\sim 100-500$ collisions)
the orbit still follows closely the nearest adiabatic curve.

Thus we propose that the 
frequency of the resonance will be well-described by the standard
semiclassical (eikonal) quantization of the integrable dynamics defined
by the adiabatic approximation, 
even though the resonance lifetime is crucially dependent
on the slow chaotic diffusion away from this curve.  With this assumption,
the resonance frequency can be calculated as a function of deformation
by a variant of the eikonal technique of ref. 14; good
agreement with exact numerical solutions is found (Fig.~2{\it b}).
The mode indices (quantized actions) are fixed to their values at
zero deformation and the calculation yields both the resonance frequency
and the adiabatic curve (value of $S$) corresponding to a given resonance.
This adiabatic curve then defines the initial conditions
for our ray calculations of the lifetime of a given resonance.  

Using the ray model for the resonance lifetime, we calculate the
mean escape time of an ensemble of rays 
launched with uniform density on the 
appropriate adiabatic curve. At each collision a
ray is allowed to escape with a transmission probability
(see Fig.~3 legend) which takes into account
both the direct tunneling which is present without deformation,
and the Fresnel scattering once $\sin \chi < 1/n$.
In Fig.\ 3 the ray prediction for the lifetime is compared to exact
wave solutions for three resonances associated with the same 
initial adiabatic curves but very different wavelengths.
Note that above the critical deformation for refractive escape 
the widths of all the resonances
agree up to factors of order unity and also agree well with the ray model.
This confirms the
universality of the broadening arising from chaotic diffusion.

\vspace*{-0.7cm}\hspace*{-2.5cm}\psfig{figure=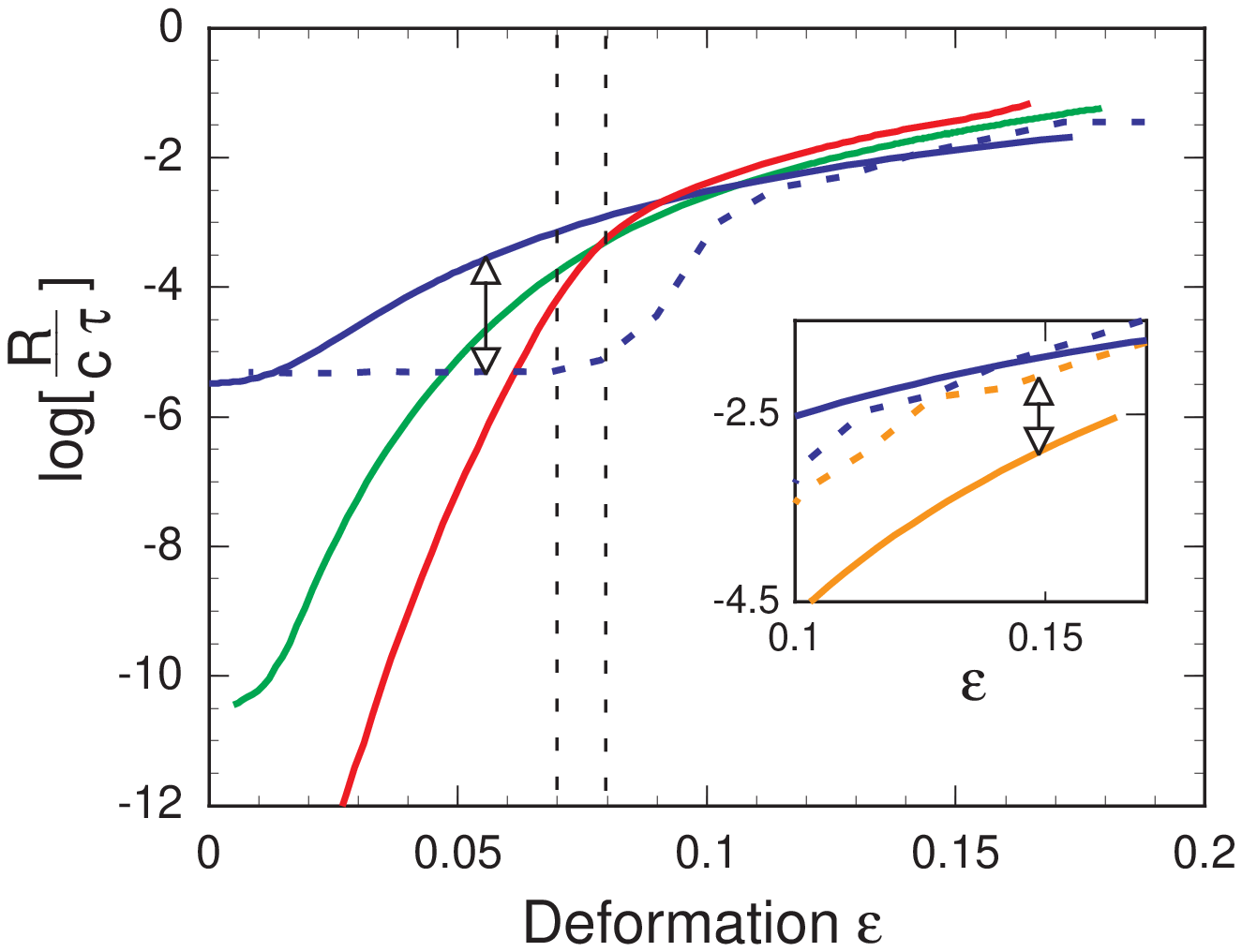,width=14cm}\vspace*{-3.5cm}
\footnotesize\noindent
FIG.~3
Logarithm of inverse resonance lifetimes (in units of $R/c$)
vs. deformation for the same resonances as in Fig. 2b 
(same color coding), determined numerically.
All three resonances correspond to an initial adiabatic curve 
with $\sin \chi_0 \approx 0.8$.
Dashed blue line is the result for this adiabatic curve 
from the ray-optics model for the $kR=12.1$ resonance.
The escape probability for collisions at $\sin \chi$ is determined
by the semiclassical relation $\sin \chi = m/nkR$ and the known
resonance lifetimes of the cylinder for angular momentum $m$
(J.U. N\"ockel et al., in preparation).
Dashed black lines delimit the cross-over region from evanescent to
refractive escape when the last intervening KAM curve breaks; as expected
all three resonances have approximately the same 
lifetime above this deformation. The discrepancy at small $\epsilon$ 
(see arrow) is indicative of chaos-assisted tunneling.
{\it Inset}: Expanded scale 
showing the resonance with $kR\approx 12.1$ and its agreement with
the ray-optics approximation (dashed blue) at high deformations.  Also shown
(brown) is a resonance with $\sin \chi_0 \approx 0.9,\,kR\approx 33.2$, 
which has a substantially longer lifetime 
than predicted by the ray-optics model (dashed brown). The difference marked 
by the arrow indicates the 
dynamical localization of photons.\\\vspace*{0.75cm}

\normalsize
However, Fig.~3 indicates two significant effects which
{\it do} depend on wavelength and are beyond the ray model.
First, we note that for small deformations, where only tunneling
escape is possible, the exact resonance
width actually increases much faster than predicted by the ray model.
We believe that this is due to chaos-assisted tunneling 
\cite{Bohigas,Doron}, since the direct (angular-momentum-conserving) tunneling 
is taken into account by the ray model.
All evanescent (tunneling) processes are forbidden within ray optics,
however the probability of such a process decreases
exponentially with the tunneling distance.  Consider the whispering
gallery orbit traversing the surface of section of Fig.~2a at $\sin \chi \approx 0.9$.
This orbit lies on an unbroken adiabatic curve and hence will never escape
according to ray optics.  However a wavepacket following this orbit has
a chance to tunnel through this dynamical barrier to the edge of 
the chaotic region at $\sin \chi = 0.8$ .
From this region, chaotic diffusion will take
the wave-packet without any further tunneling down to the critical angle
for refractive escape.  Due to the shortness of the tunneling step, 
this new escape path, which is introduced 
by the presence of chaos in a classically
inaccessible region of phase space,
strongly enhances the evanescent leakage of these resonances.

Second, we find that for resonances with the longest lifetimes
the ray model predicts too short a lifetime
at large deformations.  At large deformations refractive 
ray escape is allowed, so the longer lifetime of the wave
solution indicates that this classically-allowed escape is being
suppressed.  Precisely such an effect, known as dynamical localization,
has been studied extensively in the theory of quantum chaos.  Unlike
a particle (or ray), a wave is able to explore simultaneously 
multiple paths while undergoing chaotic diffusion.  Typically after
a characteristic time these multiple paths destructively interfere
suppressing further diffusion.  Such an effect has been observed 
indirectly through the suppression of electron ionization in atomic hydrogen
in an intense microwave electric field \cite{Casati}.

Finally, we briefly analyse the onset of directional emission shown in Fig.~4,
and discussed in more detail elsewhere \cite{OptLett2}. 
As shown in Fig.~2{\it a}, the dynamics of whispering gallery
orbits consists of a rapid motion along adiabatic curves and a slow chaotic
diffusion transverse to them.  Hence the whispering gallery orbits of interest,
which begin far from the critical line $\sin \chi = 1/n$, will diffuse slowly
until the adiabatic curve is reached which is tangent to the critical line 
and then will escape rapidly near the points of tangency (Fig.~2{\it a}).  These
points correspond to the points of maximum curvature of the ARC according
to Eq.~(1). Rays escaping near the critical angle are emitted roughly
tangent 
to the surface predicting strong emission maxima in the far-field
in directions tangent to the points of maximum curvature (Fig.~4{\it a}).
This model predicts that all ARC whispering gallery resonances
will have approximately the same directional emission pattern for a given 
index of refraction (Fig.~4{\it b}).  This universality persists
even when the adiabatic approximation breaks down, as is the case in 
Fig.~4{\it c} where the index of 
refraction is such that there are islands in the surface of section 
at the minima of the relevant adiabatic curve. Then the chaotic 
trajectories
circulate around the outside of the islands (Fig.~2{\it a}). 
In 
\twocolumn[
\hbox{{\it a}\hspace*{5.3cm} {\it b}\hspace*{6.2cm}{\it c}}
\psfig{figure=paperfigures/wave.EPSF,width=5cm}
\vspace*{-5.1cm}\hbox{\hspace*{5.3cm}
\psfig{figure=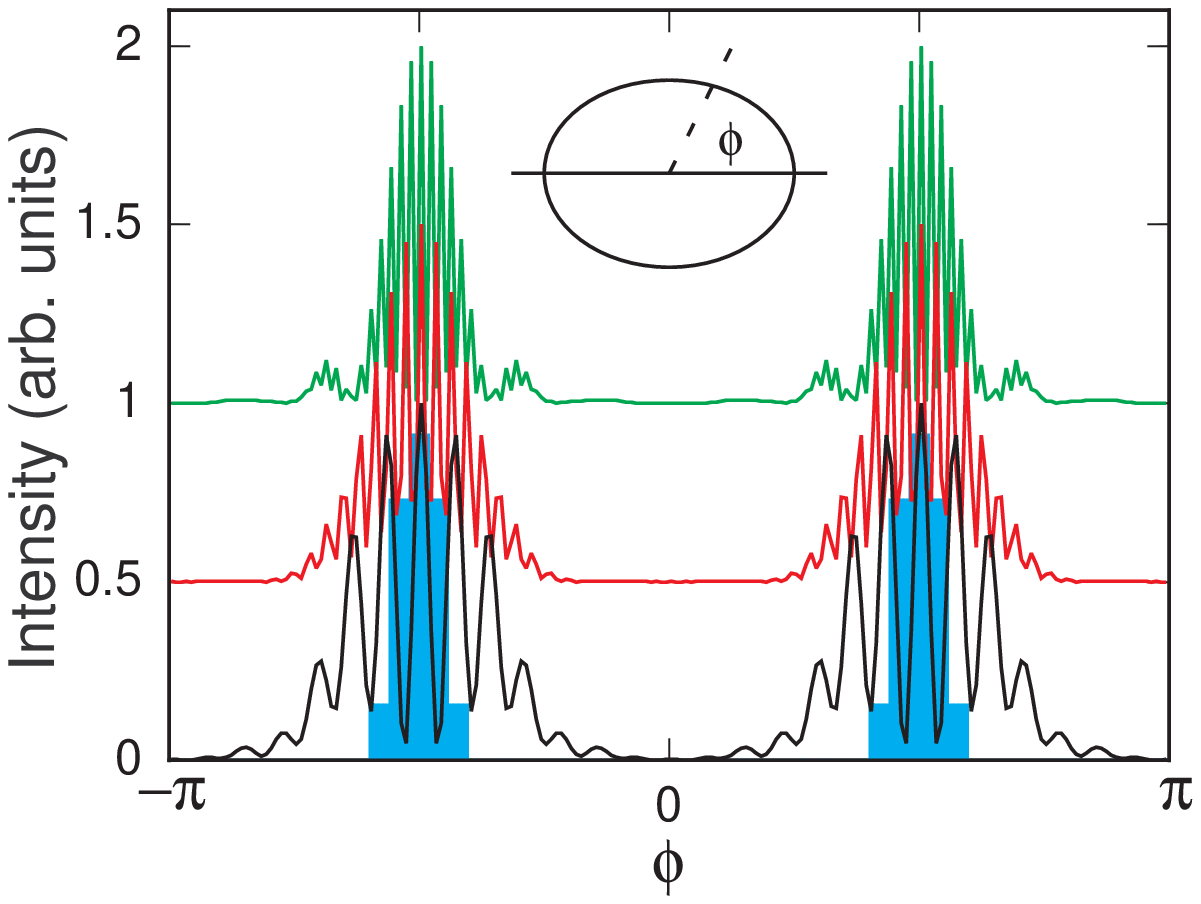,width=10cm}
\hspace*{-3.7cm}\psfig{figure=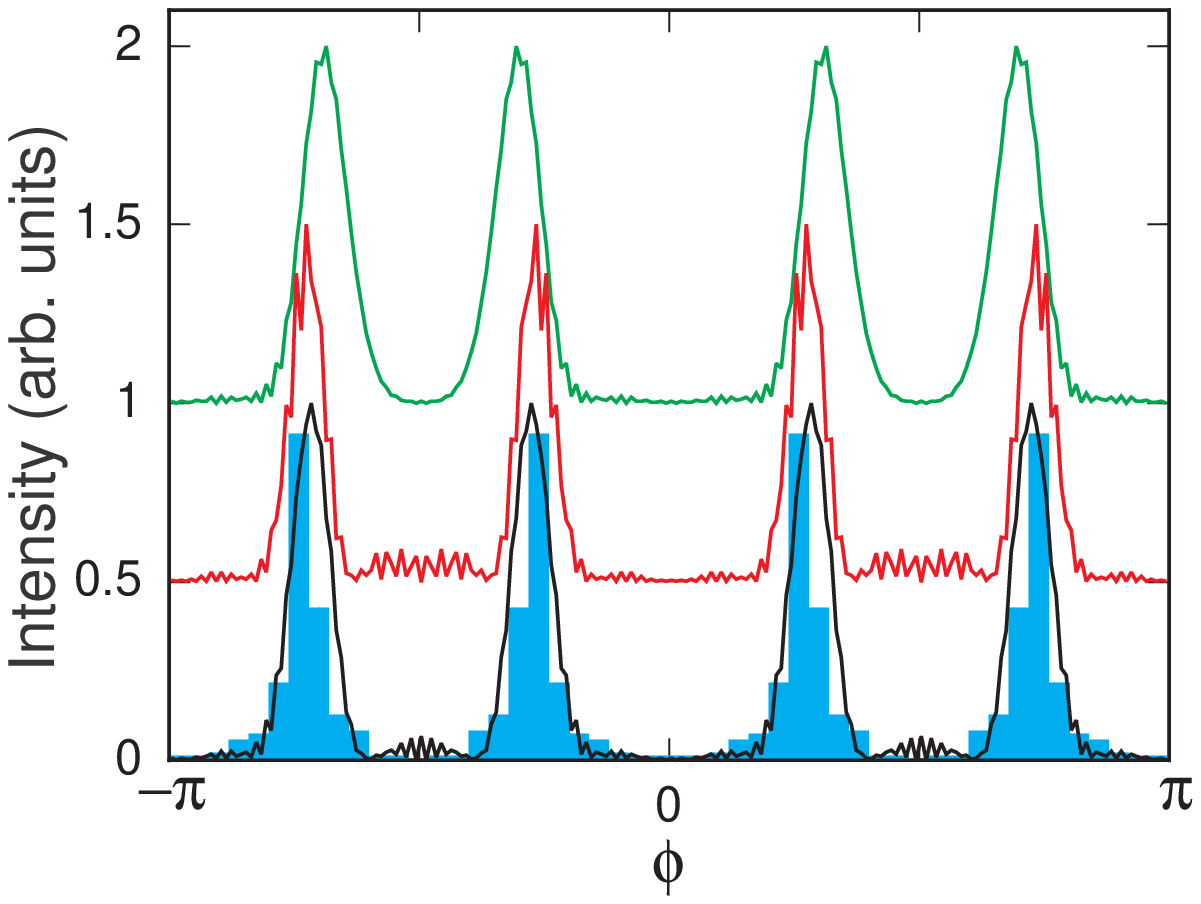,width=10cm}
}\vspace*{1cm}

\footnotesize\noindent
FIG.~4 {\it a} False color representation of the electric 
field intensity in the TM mode for the $kR=45.15$ resonance
at $\epsilon = 0.11$ determined from the
exact numerical solution.  Intensity is higher for 
redder colors, and vanishes in the dark blue regions.
One clearly sees the high intensity regions in the near-field just
outside the surface at the highest curvature points 
$\phi=0,\pi$, and the high emission intensity lines
(green) emanating from these points in the tangent directions.
{\it b} Far-field plots of the intensity of three resonances with $kR
=12.1,\sin\chi\approx 0.8$ (black); $kR=27.9,\sin\chi\approx 0.9$ (red);
$kR=45.4,\sin\chi\approx 0.75$ (green).
Index of refraction is $n=2$ (offset for clarity). Blue histogram is
the prediction of the ray-optics model.  Note two peaks at $\phi=\pm\pi/2$,
corresponding to the tangent directions to the points of 
highest curvature.
{\it c} Far-field plots for three resonances with
$kR=57.8,\sin\chi\approx 0.8$ (black);
$kR=48.4,\sin\chi\approx 0.91$ (red);
$kR=45.7,\sin\chi\approx 0.91$ (green);
compared to the ray-optics model (blue histogram) for $n=1.54$.
Note that each peak is split and the intensity
is negligible in the direction tangent to the points of highest curvature.
This arises from the presence of stable islands eclipsing the points
of highest curvature for $n=1.54$ as shown in Fig.~2a.
\vspace*{0.7cm}
]

\normalsize\noindent
this case the peaks split and maximum emission does not occur at
the points of highest curvature. All these ray-optics predictions are
seen to be in agreement with the wave solutions. 

These compact dielectric resonators with controllable Q and highly directional
emission may well be useful for applications to microlasers and
fibre-optic communications.\\
\vspace*{0.5 cm}

{\sf \noindent ACKNOWLEDGEMENTS. We thank R.~Chang, A.~Mekis,
H.~Bruus, G.~Hackenbroich and M.~Robnik for discussions.  
This work was supported 
by the US National Science Foundation and the US Army Research
Office.}\\
\vspace*{0.5 cm}

\noindent 
Received 12 July; accepted 19 November 1996

\end{document}